# *In-vitro* iontophoresis on pinneal *Sus scrofa* skin and its transport flux modeling as influenced by time and current density

## Iontoforesis pada kulit telinga babi (*Sus scrofa*) *in-vitro* dan pemodelan *transport flux*-nya yang dipengaruhi waktu dan kerapatan arus

**Tyas Pandu Fiantoro[1], Akhmad Kharis Nugroho[2]**
[1]. Electrical Engineering Undergraduate, Gadjah Mada University, eltroyaz@mail.ugm.ac.id
[2]. Faculty of Pharmacy Gadjah Mada University, Sekip Utara Yogyakarta

## Abstract

The purpose of this study was to enhance the existing time dependent flux model for the transdermal iontophoretic transport of drugs. This study evaluated the flux data as influenced by time and current density. *In vitro* iontophoresis performed on the piglet (*Sus scrofa*) necropsy–taken *medial scapha* pinneal skin that mounted in the U shaped sink chamber. Iontophoresis of atenolol with a constant dose of 1000 ppm was implemented for 3 hours with acceptor phase sampling every 30 minutes. Data were analised based on exponential fitting of each current density value to produce a current density dependent flux model. This model then combined with the time differential model of flux to produce a flux model that takes account of both current density and time.
**Key words:** modelling, flux, iontophoresis, current

## Abstrak

Penelitian ini ditujukan untuk memperbaharui model flux pada penerapan iontoforesis yang semula hanya bergantung pada waktu menjadi turut bergantung pada kerapatan arusnya. Dalam penelitian ini, data flux diamati berdasar pengaruh waktu dan kerapatan arus. Iontoforesis *in vitro* dilakukan memakai kulit pineal babi (*Sus scrofa*) bagian *medial scapha* yang dipasang pada kompartemen *U-sink chamber*. Iontophoresis atenolol dengan dosis tetap 1000 ppm diterapkan selama 3 jam dengan pengambilan sampel dilakukan setiap 30 menit. Dilakukan *fitting* fungsi eksponensial dari data yang didapat sepanjang sumbu arus, sehingga muncul model flux yang hanya bergantung pada arus. Model ini kemudian digabungkan dengan model flux yang bergantung terhadap waktu, menghasilkan sebuah model flux yang bergantung pada kerapatan arus dan waktu.
**Kata kunci:** pemodelan, fluks, iontoforesis, arus

## Introduction

The simple time based iontophoretic models made by Nugroho *et al.* in 2005 proved to be useful for predicting the administered dose of iontophoretic drug delivery. The model, however, limited to a constant current density value, hence, to modify the transport profile, altering the donor phase concentration, duration of the iontophoresis, or even the applied surface area must be done. This article proposes an enhanced model of the iontophoretic transport flux based on *in-vitro* experiment. Resulting model hopefully allows the convenient flux profile alteration by changing its current density. Hence controlling iontophoretic drug dose will be as ease as controlling electrical current.

## Methodology
### Materials
Atenolol (4-(2-hidroxy-3-[(1-methylethyl) amino]propoxy benzenacetamide) with 99,5% purity, obtained from Calao, Milan, Italy. Propylene glycol, oleic acid, mannitol, di-sodium hydrogen phosphate (Na2HPO4), Sodium di-hydrogen phosphate (NaH2PO4) and sodium chloride (analytical grade, Merck, Darmstadt, Germany) were purchased from PT General Labora Yogyakarta Indonesia. Whereas Sodium chloride (NaCl), potassium chloride (KCl), tri-sodium citrate di-hydrate ($Na_3C_6H_5O_7.2H_2O$), citric acid anhydrate ($C_6H_8O_7$), aquadest ($H_2O$), Calcium oxide (CaO), ethanol 90% ($C_2H_5OH$), and pH indicator paper (MERCK), provided by the Laboratory of Biopharmaceutics, Unit III, Pharmacy Faculty, Gadjah Mada University (UGM). Current was measured using HELES UX 839-TR multimeter. Digital scale Sartorius BP 310P, digital caliper Mitutoyo Digimatic, clay surgery bed, magnetic stirrer Stuart, and GENESYS 10S spectrometer (software version v1.200 2L9K274001, Thermo Scientific) are also provided by Laboratory of Biopharmaceutics, Unit III, Pharmacy Faculty, Gadjah Mada University (UGM). U shaped sink chambers used are properties of A .K. Nugroho. The used iontophoretic current regulator device was capable for administering a dose of 0 μA, 378 μA, 731 μA, or 1130 μA, each with maximum 3% tolerance.

**Skin Preparation**

The used piglet skin comes from the *medial scapha* region rather than the *lateral scapha* region due to its epidermal thickness. Approximate surface area of three cm$^2$ of *medial scapha* skin was extracted using a scalpel, then rinsed once with API (*aquadest pro injection*) before soaked in 50 mL solution of 0.15 M phosphate buffer saline at pH 7.4 within 15 minutes. Afterwards, the treated skin placed on the acceptor chamber mounter with a diameter of 11.86 mm as illustrated by Fig. 1. The inner area of the mounted skin is always 1.1 cm$^2$. Excess skin on the outer area then cut away with a cutter. After a proper mounting of skin, the cell was then secured by joining and screwing the donor and the acceptor parts together as shown in Fig. 2.

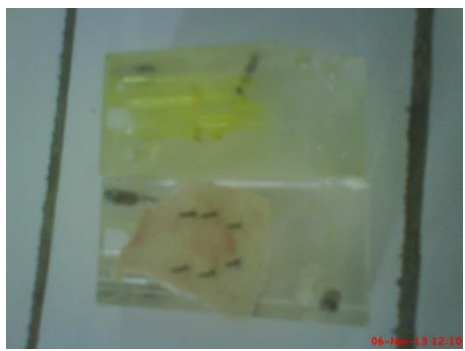

Fig. 1. The mounted piglet skin

**In-vitro Iontophoretic Studies**

The donor compartment was then filled in with 1000 ppm atenolol solution (in 5mM citric buffer at pH of 5). The acceptor phase was filled with 2 ml of 0.15M phosphate buffer saline at pH 7.4. Mannitol was added (6 g in 250ml buffer) in the donor solution to balance the tonicity of donor and acceptor media (Nugroho, 2011). Iontophoretic transport study was performed for 3 hours, with the U shaped sink chambers placed on the magnetic stirrer as shown in Fig. 3.

Every 30 minutes, 1.5 ml samples were collected periodically from the acceptor compartment. Each sample's 216 nm wavelength absorbance then read using UV-Vis spectrometry. The calibration curve has r > 0.94 in the concentration range of 0.2 ppm to 20 ppm of atenolol. Because the acceptor compartment volume is 2.5 ml, acceptor phase amount is only 2 ml. After each sampling, 1.5 ml of 0.15 M phosphate buffer saline at pH 7.4 was added to the acceptor chamber.

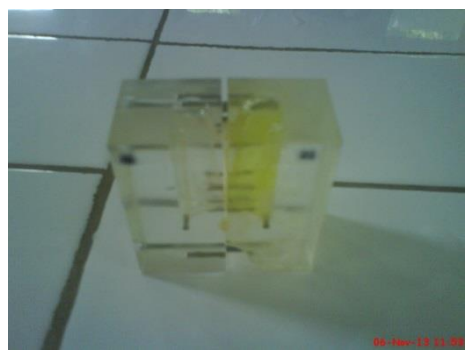

Fig. 2. The secured U shaped sink chambers

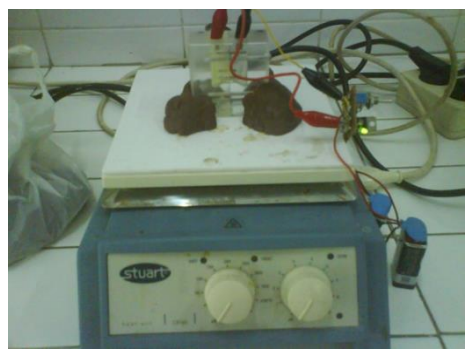

Fig. 3. The magnetic stirrer, turned off

The transported atenolol mass for each sample is calculated according to (1). *A* is the read absorbance value for each sample number (indexed by *p* or *q*). *V* is the sample volume which happened to be always 1.5 ml for all index (both *p* and *q*).

$$m_p = \frac{4}{3}\left(V_p f(A_p) - \sum_{q=1}^{p-1} \frac{1}{4^p} V_q f(A_q)\right) \quad (1)$$

The *f* function used in (1), described in (2). This is the Beer-Lambert's reverse function to determine the concentration of the mixture, based on the calibration curve.

$$f(A) = \frac{A + 0.0188}{0.1411} \quad (2)$$

The accumulation of the atenolol mass transported is a mere sum of each previous transported mass as given in. Hence, the total transported atenolol after three hours (that is six time of samplings) of iontophoresis is given in (3). Therefore, the flux per 30 minutes is obtained by (4), where $S$ is the mounted skin's surface area.

$$M = \sum_{p=1}^{6} m_p \qquad (3)$$

$$J_p = \frac{m_p - m_{p-1}}{S} \qquad (4)$$

**Results**

The flux observed in (4) could be expressed as (5) for the modelling its continuous process. Nugroho *et al.* in 2005 propose a model shown in (6) that could be fitted nicely with the obtained data. $M$ is the continuous transported atenolol mass function. Substituting (6) into (5) and then by solving the resulting differential equation, solution (7) is obtained.

$$J = \frac{\partial M(t)}{S\, \partial t} \qquad (5)$$

$$\frac{\partial M(t)}{\partial t} = i_0 - K_R\, M(t) \qquad (6)$$

$$J = \frac{i_0}{S}(1 - e^{-K_R t}) \qquad (7)$$

It was observed that the flux did not linear with its current density for the applied current range of 0 μA to 1130 μA. Instead, the exponential model expressed in (8) is more suited with the observed flux data.

$$\frac{\partial J}{\partial i} = a\, e^{b-i} \qquad (8)$$

The flux function, $J$ modeled as separable by multiplication, as expressed by (9). Substituting each of (7) and (8) to (9), equation (10) was formed. Then, by taking the integrals, the solution of $J$ is obtained as described by (11).

$$\frac{\partial^2 J}{\partial i\, \partial t} = \frac{\partial J}{\partial i}\, \frac{\partial J}{\partial t} \qquad (9)$$

$$J = \int a\, e^{b-i}\, \partial i \int \frac{\frac{i_0}{S}(1 - e^{-K_R t})}{\partial x}\, \partial x \qquad (10)$$

$$J = \left(C_0 - \frac{a}{e^{i-b}}\right)\frac{i_0}{S}(1 - e^{-K_R t}) \qquad (11)$$

With the initial value of $J = 0$ when $i = 0$, and final $J$ value of $d$ for $i = \infty$, the value of $C_0$ is determined as expressed in (12). Substituting (12) to (11) yielded (13). Each $a$, $b$, $c$, and $d$ constant could be simplified with (14), (15), and (16), yielding the simpler equation (17).

$$C_0 = d\, e^{K_2} \qquad (12)$$

$$J = \frac{i_0}{a\, S}\left(\frac{d}{a} e^{K_2} - \frac{1}{e^{i-b}}\right)(1 - e^{-K_R t}) \qquad (13)$$

$$K_0 = \frac{i_0}{a} \qquad (14)$$

$$K_1 = \frac{d}{a} \qquad (15)$$

$$K_3 = b \qquad (16)$$

$$J = \frac{K_0}{S}\left(K_1 e^{K_2} - \frac{1}{e^{i-K_3}}\right)(1 - e^{-K_R t}) \qquad (17)$$

The final flux equation stated in (17) could be used as a basis for both iontophoretic and post-iontophoretic transport flux modelling. Since the two states basically a different events, it is best to model the transport flux for iontophoretic and post-iontophoretic period with a piecewise functions (Nugroho *et al.*, 2005). Now the model could be expressed with the dependency of both time and current density variable, as stated by (18).

$$J = \begin{cases} \varepsilon\,(1 - e^{-K_R t}) & ;\, t < t_0 \\ \varepsilon\, P_{PI}(1 - e^{-K_R t}) + \varepsilon\, K_R\, X_T\, e^{-K_R t}; & t \geq t_0 \end{cases}$$
$$(18)$$

For the sake of brevity, an abbreviation of ε, which described in (19) was used.

$$\varepsilon = \frac{K_0}{S}\left(K_1 e^{K_2} - \frac{1}{e^{i-K_3}}\right) \qquad (19)$$

The full three dimensional plot of equation (18) could be seen at Fig. 4.

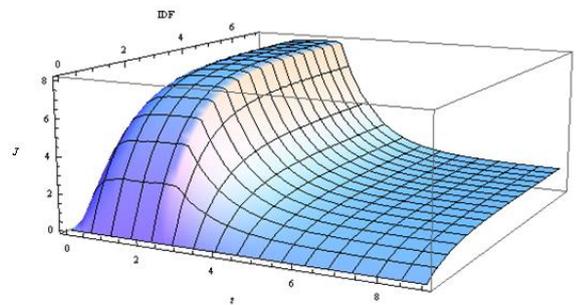

Fig. 4. The resulted full model